\documentclass[aps,prl,nofootinbib,twocolumn]{revtex4-1}

\usepackage{epsfig,amssymb,amsmath}

\usepackage{color}
\usepackage[utf8]{inputenc}
\usepackage{amsmath, amssymb, mathtools, braket, graphicx, float, bbold, bm}
\usepackage{hyperref}
\newcommand{\mods}[1]{\left|#1\right|^2}
\hypersetup{
    colorlinks=true,
    linkcolor=blue
}

\usepackage[colorinlistoftodos]{todonotes}
\usepackage{soul}

\def\labell#1{\label{#1}}
\def\>{\rangle}\def\<{\langle}
\def\section#1{{\par\em #1:--- }}
\def\togli#1{}

\newcommand{\Mod}[1]{\left|#1\right|}

\begin{document}


\title{Quantum computers to test fundamental physics or viceversa }

\author{Simanraj Sadana$^1$, Lorenzo Maccone$^2$, Urbasi Sinha$^{1}$ }
\email{usinha@rri.res.in}
\affiliation{ \vbox{$^1$Light and Matter Physics, Raman Research Institute, Bengaluru-560080, India}
\vbox{$^2$Dip.~Fisica and INFN
    Sez.~Pavia, University~of Pavia, via Bassi 6, I-27100 Pavia,
    Italy}}
\begin{abstract}
  We present two complementary viewpoints for combining quantum
  computers and the foundations of quantum mechanics. On one hand,
  ideal devices can be used as testbeds for experimental tests of the
  foundations of quantum mechanics: we provide algorithms for the
  Peres test of the superposition principle and the Sorkin test of
  Born's rule. On the other hand, noisy intermediate-scale quantum
  (NISQ) devices can be benchmarked using these same tests. These are
  deep-quantum benchmarks based on the foundations of quantum theory
  itself. We present test data from Rigetti hardware.
\end{abstract}


\maketitle  
Physics is experimental, so the postulates of all physical theories
are based on experiments. Up to now, the role of quantum computers in
fundamental physics has mainly been limited to the simulation of
complex quantum systems \cite{feynman}.  Here, instead, we propose to
use them directly for experimental tests of the postulates of quantum
mechanics. In the ideal case, assuming perfect hardware, they are
especially suited to this aim as they are quantum systems with a large
number of degrees of freedom. In contrast, in the non-ideal case of
NISQ devices, one can assume that quantum mechanics is valid and use
these tests for fundamentally benchmarking the device, since they are
based on the very foundations (the postulates) of the theory. In other
words: assuming perfect hardware, one can test quantum mechanics;
assuming quantum mechanics, one can test the hardware. Relaxing both
assumptions, one can perform self-consistency checks to test both.

We present two such experimental tests: we give algorithms and quantum
machine code for the Peres and the Sorkin tests and run them on
Rigetti quantum computers. The first one is a test of the state
postulate of quantum mechanics (i.e.~the superposition principle),
which claims that quantum states live in a {\em complex} Hilbert
space. In principle, one could imagine a quantum mechanics based on
real \cite{woottersreal,stu}, complex, or quaternionic Hilbert spaces
\cite{perescompl}: the choice is based on the outcome of experiments,
such as the Peres one, see also
\cite{toni,walther,greg,gregor,janwei,imagin}. The fact that complex
numbers are necessary (and sufficient) has interesting implications,
e.g.~it implies that quantum states are locally discriminable
\cite{paoloreconstruction} and it is connected to the locality of some
quantum phenomena \cite{toni}. The second one, proposed by Sorkin
\cite{sorkin}, is a test of the Born postulate. The Born rule declares
that quantum probability is the square modulus of a scalar product in
the state space. A failure (or an extension \cite{caslav}) of the Born
rule would result in a new physical effect: the presence of genuinely
$n$-fold superpositions that cannot be reduced to an iteration of the
usual two-fold superpositions we find in textbook quantum mechanics
\cite{caslav,zykowski,lee}.
Thanks to our implementation, we also ran
both tests at the same time for a class of states. In contrast to
previous tests
\cite{urbasi,laflamme,werner,walther,greg}, ours do not use custom-built setups, they permit
arbitrary initial states, and can be easily scaled up as new reliable
quantum computers become available.

There are multiple advantages of doing these (and other) fundamental
tests on quantum computers: (i)~Both tests are performed on the same
hardware, which prevents possible biases that may arise from a
tailored experimental setup. At the same time, it is simple to
translate the proposed algorithms to different quantum computer
architectures, if one wants to confirm the results independently.
(ii)~It is possible to perform both experiments {\em at the same time}
(see below), which is important since, as discussed below, the two
experiments are not entirely independent of each other. (iii)~These
experiments are easily scalable to large dimensions, once reliable
quantum computers are available. (iv)~As discussed one can reverse
perspective: under the assumption that quantum mechanics is correct,
these tests become deep benchmarks for a quantum computer.
\section{Peres test}\label{sec:peres_test_theory}
The state postulate claims: ``The pure state of a system is described
by a normalized vector $|{\psi}\>$ in a complex Hilbert space''. As
all physical postulates, it is based on experimental data, and the
Peres' test specifically refers to whether one needs complex numbers,
reals \cite{stu}, quaternions \cite{perescompl,zykowski,adler},
octonions \cite{lee,octonions}, etc., but it does not question the Hilbert
space structure of the theory. For example, we accept the natural
assumption that the Hilbert space dimension is equal to the system's
number of degrees of freedom, namely of independent outcomes of a
nondegenerate observable (dropping this assumption
\cite{woottersreal,stu}, one needs different tests for the complexity
of quantum mechanics \cite{toni,greg,walther,janwei}, based on the
locality of measurement outcomes). Octonions can be discarded upon
observing that they are not associative for multiplication
(interestingly, this means that different combinations of two-fold
interferences may give different results, which would give the same
signature as a failure of the Sorkin test).

We now review the Peres test \cite{perescompl}.  Consider two pure
states $\ket{\psi_1}$ and $\ket{\psi_2}$, and their superpositions
$\ket{\psi_{12}} = \alpha \ket{\psi_1} + \beta \ket{\psi_2}$ with
$\alpha,\beta$ nonzero reals. If we project it on, say, $\ket{1}$ (any
other state would give similar results), then, assuming complex
Hilbert spaces, the probability of successful projection is
\begin{align}
  \mods{\braket{1|\psi_{12}}} 
  =& \mods{\alpha \braket{1|\psi_1}} + \mods{\beta \braket{1|\psi_2}} \nonumber \\
   &+ 2 \alpha \beta \left|\braket{1|\psi_1} \braket{1|\psi_2}\right| \cos\varphi_{12} \\
    \label{eq:cosphi12}
  \implies\cos\varphi_{12} =& \frac{\mods{\braket{1|\psi_{12}}} - \mods{\alpha \braket{1|\psi_1}} - \mods{\beta \braket{1|\psi_2}}}{2 \alpha \beta \left|\braket{1|\psi_1} \braket{1|\psi_2}\right|}
\end{align}
with
$\varphi_{12} = \arg(\braket{1|\psi_1}
  \braket{\psi_2|1})$. If, instead we assume that real
Hilbert spaces are sufficient, the term $\cos\varphi_{12}$ can only
take the values $\pm 1$. We can rewrite the left-hand-side of
\eqref{eq:cosphi12} in terms of experimental values as
\begin{align}
\label{eq:gamma_def}
  \gamma_{12} \coloneqq \frac{p_{12} - \alpha^2 p_1 - \beta^2p_2}{2 \alpha \beta\sqrt{p_1p_2}}
\end{align}
with $p_{12}$, $p_1$ and $p_2$ the experimental probabilities of
projection onto $|1\>$ of $|\psi_{12}\>$, $|\psi_1\>$ and $|\psi_2\>$.
If we experimentally find that $\gamma_{12}=\pm 1$ {\em always}, then
a real quantum theory is sufficient. If we find states for which
$|\gamma_{12}|<1$, then it is necessary to use a complex or
quaternionic quantum theory.  If $|\gamma|>1$, the superposition
principle is violated.

To discriminate between a complex and a quaternionic theory we need a
further step, based on the identity
\begin{align}
\label{cosid}
    \cos^2a + \cos^2b + \cos^2c - 2 \cos a\cos b \cos c = 1
\end{align}
valid for any $a,b,c$ real numbers with $a + b + c = 0$.  Consider
three pure states $\ket{\psi_1}$, $\ket{\psi_2}$, $\ket{\psi_3}$, and
take superpositions of two at a time. We will have three quantities of
the type given in \eqref{eq:cosphi12} with $\varphi_{12}$,
$\varphi_{23},$ or $\varphi_{31}$, one for each pair.  Since
$ \varphi_{12} + \varphi_{23} + \varphi_{31} = 0$, then the identity
\eqref{cosid} holds for these three angles {\em if} a complex quantum
theory is sufficient. Otherwise, if a quaternionic theory is
necessary, the amplitudes cannot be represented by vectors in a 2D
plane and therefore the LHS in \eqref{cosid} is less than 1 in
general. One can detect this by analyzing the quantity
$F = \gamma_{12}^2 + \gamma_{23}^2 + \gamma_{31}^2 - 2
\gamma_{12}\gamma_{23}\gamma_{31}$ (where $\gamma_{23}, \gamma_{31}$
are defined similarly to $\gamma_{12}$, but using the state
$\ket{\psi_3}$). If the experimentally measured $F$ is {\em always}
one for all states, then a complex theory is sufficient (no
quaternions, octonions, etc.~are needed) since all the $\gamma$s can
be written as cosines, and \eqref{cosid} holds. Otherwise if
$|F| < 1$, then we must employ quarternions.  Finally, if $|F| > 1$,
the superposition principle is violated, as the states
  cannot be represented as vectors.

\begin{figure}[H]
    \centering
    \includegraphics[width=0.6\hsize]{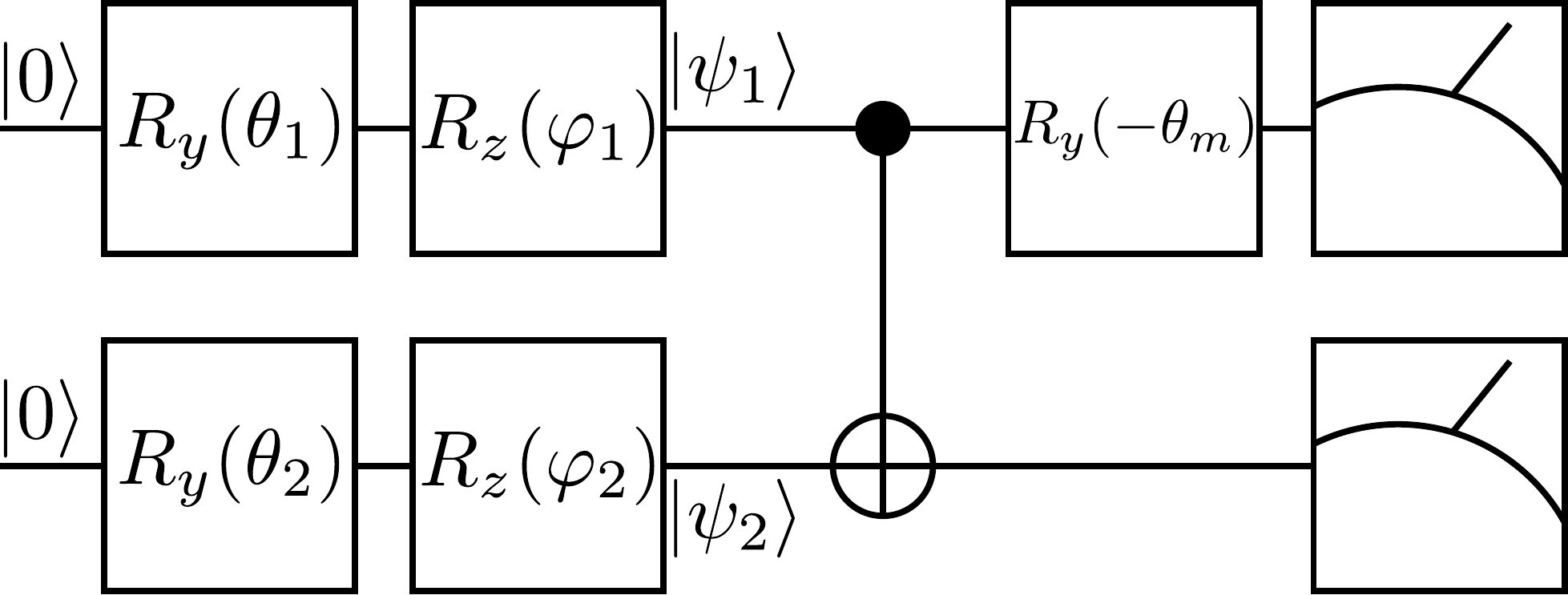}
    \caption{Graphical depiction of the algorithm to calculate the
      $\gamma$ factor of \eqref{lll} by creating and projecting
      superpositions of arbitrary states. The states are created from
      unitary rotations $R_\ell(\lambda)$ of an angle $\lambda$ around
      the $\ell$ axis, with $a_k=\cos\theta_k/2$,
      $b_k=\sin\theta_k/2$, followed by a CNOT and a rotation by angle
      $-\theta_m$ for desired projection. For example, to project on
      $\alpha\ket{01} + \beta\ket{10}$, choose
      $\theta_m = 2\cos^{-1}\alpha$ and record the counts of
      $\ket{01}$. The unentangled projections
      $|\braket{01|\psi_1\psi_2}|^2$, etc., that are necessary for the
      $\gamma$s can be measured by a similar circuit where the
      CNOT-gate and the single-qubit rotation at the end are removed.}
    \label{f:peres_circ}
\end{figure}

The above Peres proposal can be directly implemented on a quantum
computer using a unary encoding (one qubit per system) where
orthogonal states are mapped into separate physical qubits and their
superpositions are obtained through interferences among them, but our
tests showed that such procedure is highly sensitive to noise and will
be reported elsewhere \cite{simanraj}.  Moreover, one has to make sure
that the algorithm does not contain as input the quantity
$\cos\varphi_{12}$ of \eqref{eq:cosphi12}, which would render the
whole procedure circular. We now present a nontrivial way to overcome
both problems: it is suited to current NISQ devices and the cosine
term only arises from quantum interference of different paths.

The trick is to prepare a two-qubit factorized state
$\ket{\psi_1}\ket{\psi_2}$ with
$\ket{\psi_k}=a_k\ket{0}+e^{i\varphi_k}b_k\ket{1}$ ($a_k\in[0,1]$,
$b_k=\sqrt{1-a_k^2}$), and then project it onto the anti-correlated
subspace spanned by $\ket{01}$ and $\ket{10}$. This produces a state
proportional to \begin{align}
&a_1b_2e^{i\varphi_2}\ket{01}+a_2b_1e^{i\varphi_1}\ket{10}\propto
\labell{anticorr}
a_1b_2e^{i\varphi_2}(\alpha^*\ket{\Psi^+}+\;\\&
\beta\ket{\Psi^-})+
a_2b_1e^{i\varphi_1}(\beta^*\ket{\Psi^+}-\alpha\ket{\Psi^-}),
\labell{psip}
\\&
\mbox{with }
\ket{\Psi^+}={\alpha\ket{01}+\beta\ket{10}},\ 
\ket{\Psi^-}={\beta^*\ket{01}-\alpha^*\ket{10}}.\nonumber
\end{align}
By projecting this state onto $\ket{\Psi^+}$, we can see the interference
among the two $\ket{\Psi^+}$ paths present in the state \eqref{psip}.
Indeed,
\begin{eqnarray}
  &&2\Mod{\braket{\Psi^+|\psi_1\psi_2}}^2 =
 \Mod{\alpha}^2\Mod{\braket{01|\psi_1\psi_2}}^2 +
     \Mod{\beta}^2\Mod{\braket{10|\psi_1\psi_2}}^2 +
\nonumber\\&&\label{lll}
 2 \mathrm{Re}\left\{\alpha^*\beta\right\}\Mod{\braket{01|\psi_1\psi_2}\braket{10|\psi_1\psi_2}} \cos\varphi_{12}  \\
&&    \Rightarrow
 \cos\varphi_{12} = \tfrac{2\Mod{\braket{\Psi^+|\psi_1 \psi_2}}^2 -
   \Mod{\alpha}^2\Mod{\braket{01|\psi_1 \psi_2}}^2 -
   \Mod{\beta}^2\Mod{\braket{10|\psi_1
   \psi_2}}^2}{2\mathrm{Re}\left\{\alpha^*\beta\right\}
   \Mod{\braket{01|\psi_1 \psi_2}\braket{10|\psi_1 \psi_2}}}
   \nonumber 
\end{eqnarray}
Projections onto an
entangled state can be implemented by a CNOT-gate and a single-qubit
rotation followed by a measurement in the computational basis. The experimental values of the
$\gamma$s can then be obtained by measuring the probability of
projection of this state (and of the projection of $\ket{\psi_1\psi_2}$
onto $\ket{01}$ and $\ket{10}$). The algorithm to create and measure these
states is given pictorially in Fig.~\ref{f:peres_circ}. Once the
$\gamma$'s are measured, we can test, by hypothesis testing, whether
their experimental values are compatible with $\pm 1$.  Similarly for
$F$.  In principle, the Peres test should check that $F=1$ for {\em
  all} states, which is, of course, not feasible. But, by choosing sets
of (uniform) random states, we sample the Hilbert space uniformly.

\begin{figure}
   \centering
\includegraphics[width=.7\hsize]{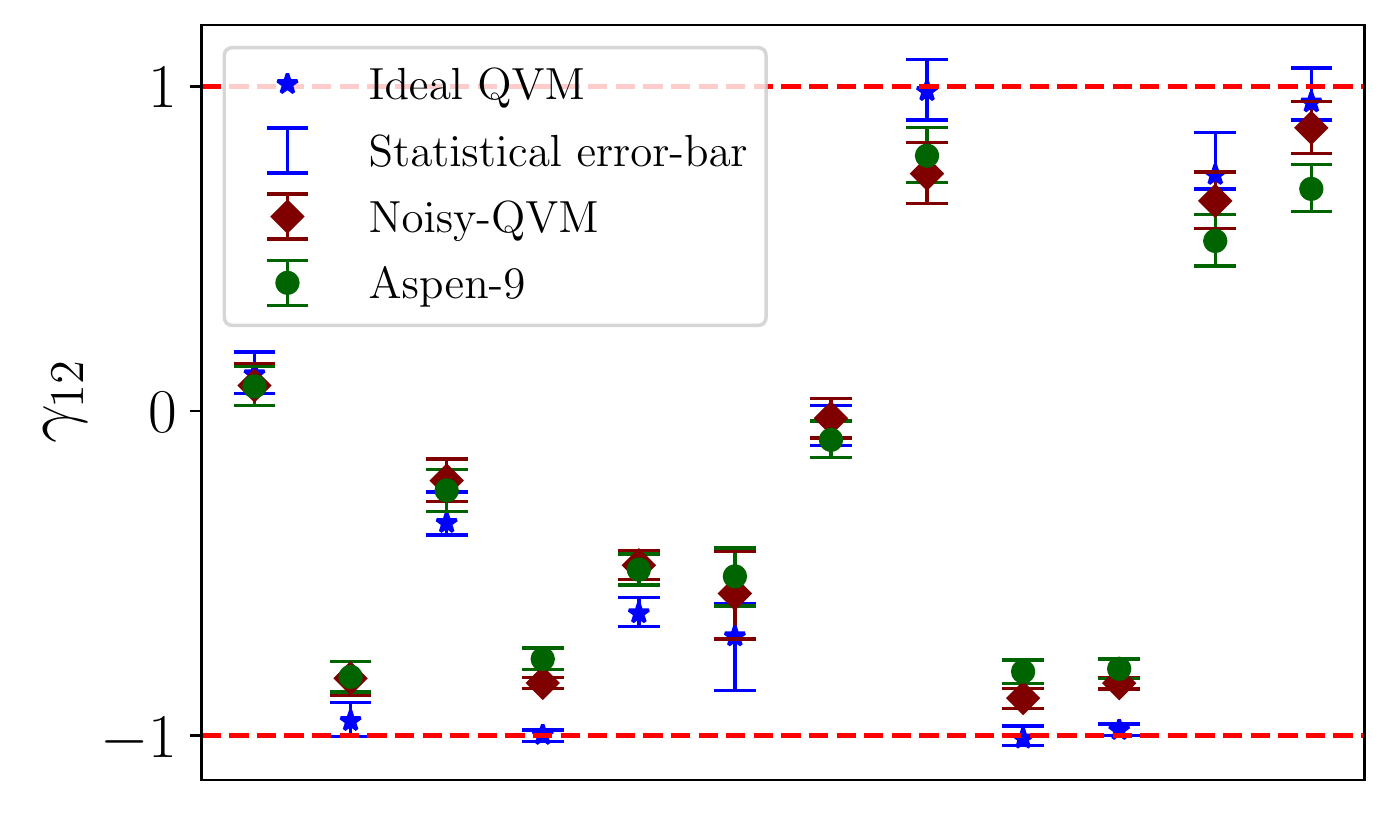}
\includegraphics[width=.7\hsize]{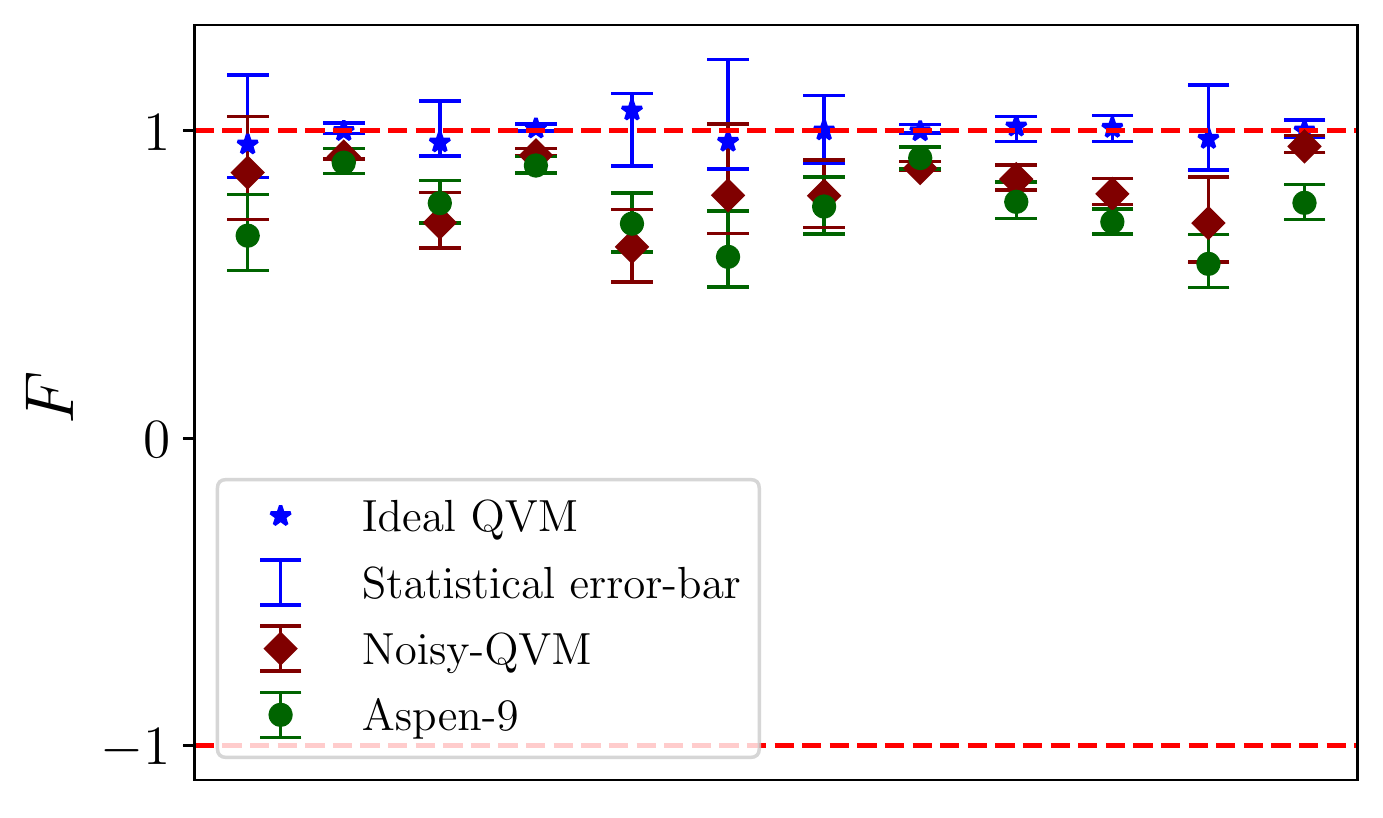}
\caption{Plots of the $\gamma_{12}$ and of $F$ for different sets of
  (uniformly distributed) random states.  The green circles are the
  measured outputs on Rigetti Aspen-9 device, the blue stars are the
  (ideal) quantum prediction from Rigetti's quantum virtual machine
  (QVM), the red squares are the quantum prediction that takes into
  account the systematic noise sources (the noise parameters are
  obtained from Rigetti's own calibration data). Most $\gamma$ values
  are different from $\pm 1$, which implies that a real Hilbert space
  is insufficient. The values of $F$ are all smaller than one, but
  they are all statistically compatible with the value corrected by
  taking into account the readout and dephasing errors in the hardware  (the green and red
  error bars overlap). Thus, the $F$ data should be considered as a
  deep-quantum benchmark for the hardware.  No quantities are outside
  the interval $[-1,1]$, so the superposition principle is not
  violated.
}
    \label{f:res_peres}
\end{figure}

The experimental results are presented in Fig.~\ref{f:res_peres}. The
source code of our algorithm using the Rigetti SDK \cite{rigettiweb}
has been uploaded in \cite{sourcecode}. The fact that a
complex quantum theory is necessary, and it is also sufficient as
quaternions are not required, is confirmed by our results up to
experimental error (which we fully characterized). All the circuits
are run for $10^4$ shots and the limited sample gives rise to
significant statistical fluctuations in the results,
  indicated in the plots by 3$\sigma$ confidence intervals. We also
  plotted the values obtained from simulations on Rigetti's Quantum Virtual Machine (QVM).
Because of noises in the quantum computer, the result of the
Peres-test has significant deviations from the theoretical values.
However, when we take into account the dominant errors (readout and
dephasing errors here) in the gates in a noisy simulation, we observe
similar deviations in the results. The samples collected from the
quantum computer and the noisy simulator are also bootstrapped to find
a confidence interval based on that sample. This confirms that the
observed value of $F<1$ is due to the noises, and we cannot reject the
hypothesis that complex numbers are sufficient. We explore the effects
of different types of errors in \cite{simanraj}.

\section{Sorkin test}
The measurement postulate (Born rule) claims: ``The probability that a
measurement of a property $O$, described by the operator with spectral
decomposition $\sum_{j}o_j\ket{j}\bra{j}$, returns a value $o_j$, given
that the system is in state $\ket{\psi}$, is
$p(o_j|\psi)={|\braket{\psi|j}|^2=|x_{j}|^2}$, with
$\ket{\psi}=\sum_{j}x_{j}\ket{j}$'' \footnote{Using Naimark's theorem, this
  formulation encompasses also measurements described by Positive
  Operator-Valued Measures (POVMs). It can be extended trivially to
  nondegenerate observables by adding a degeneracy index:
  $O=\sum_{jk}o_j\ket{j,k}\bra{j,k}$ gives a probability
  $p(o_j|\psi)=\sum_k|\braket{\psi|j,k}|^2$  \cite{luders}. }.

The linearity of quantum mechanics implies that, if a value $o_j$ of
some system property is determined by two or more indistinguishable
pathways, the probability of measuring such value is obtained from the
sum (interference) of the amplitudes for each (superposition
principle). This interference is encoded in the postulate by the
scalar product $\braket{\psi|j}$ (whose definition contains a sum). The
exponent 2 in the probability postulate implies that the superposition
of more than two pathways gives the same probability that is obtained
by separately considering the interference of all the {\em couples} of
paths independently \cite{sorkin,caslav}. Namely no {\em genuinely
  $n$-path} interference effects appear for $n>2$. In fact, for $n=3$,
assuming the Born rule, consider the following probabilities
\begin{align}
&   p_{123}=\mods{\left(\frac{\bra{1} + \bra{2} +
  \bra{3}}{\sqrt{3}}\right) \ket{\psi}} 
=\frac{\mods{x_1 + x_2 + x_3}}3 \label{tre} \\&
p_{12}=   \mods{\left(\frac{\bra{1} + \bra{2}}{\sqrt{2}}\right)
  \ket{\psi}} =\frac{\mods{x_1 + x_2}}2, \mbox{ etc.} \label{due} \\&
p_1=   \mods{\braket{1|\psi}} = \mods{x_1}, \mbox{ etc.},\label{probs}
\end{align}
where the bras refer to the $O$ eigenstates $\ket{j}$, and the term
\eqref{tre} refers to a three-path interference, \eqref{due} to
two-path interference and \eqref{probs} is the probability of each
path by itself. The exponent 2 ensures that the multipath probability
can always be expressed in terms of the two-path and single-path
ones. Indeed the quantity
\begin{align}
\kappa_3=3p_{123}-2(p_{12}+p_{23}+p_{13})+p_1+p_2+p_3
\labell{kappa3}\;
\end{align}
is identically null thanks to the following identity, valid for any
three complex numbers $x_1,x_2,x_3$:
\begin{align}
\label{eq:3_identity}
&    3\mods{\tfrac{x_1 + x_2 + x_3}{\sqrt{3}}} - 2\Big[\mods{\tfrac
  {x_1+x_2}{\sqrt{2}}} + \mods{\tfrac{x_1 + x_3}{\sqrt{2}}} +
  \mods{\tfrac{x_2+x_3}{\sqrt{2}}}\Big]\nonumber\\&
+ \mods{x_1} + \mods{x_2} + \mods{x_3} = 0.
\end{align}
Sorkin \cite{sorkin} proposed to check the form of the Born rule and
the superposition principle by {\em measuring} the probabilities
\eqref{tre}-\eqref{probs} and calculating the experimental value of
$\kappa_3$ to check if it is null (up to statistical errors). This can
be extended to arbitrary $n$. In fact, if we assume (or measure
experimentally) that the $\kappa_j$'s up to $j=n-1$ are null, one can
show by induction that
\begin{align}
\label{kappa_n}
    \kappa_n =& \Big|\sum\limits_{j=1}^{n}x_j\Big|^2 - \sum\limits_{j,k>j}^{n-1,n}|x_j+x_k|^2 + (n-2)\sum\limits_{j=1}^{n}|x_j|^2,
\end{align}
so that one can incrementally increase $n$ by just measuring the
$n$-path, the $2$-path and $1$-path probabilities.

Importantly, one has to ensure that the pathways are distinguishable
(i.e.~they are described by orthogonal states), otherwise
interferences are not obtained through simple sums as in
\eqref{tre}-\eqref{probs}. Initial experiments were carried out following Sorkin's
proposal of multi-slit experiments \cite{urbasi} which are only
approximately orthogonal (looping paths that go through multiple slits
exist \cite{urbasi2, rengaraj, usinha3}). As we do here, some tests used orthogonal
states \cite{gsch}, where a null result is easier to
evaluate.

\begin{figure}[ht]
\begin{center}
(a) \includegraphics[width=.8\hsize]{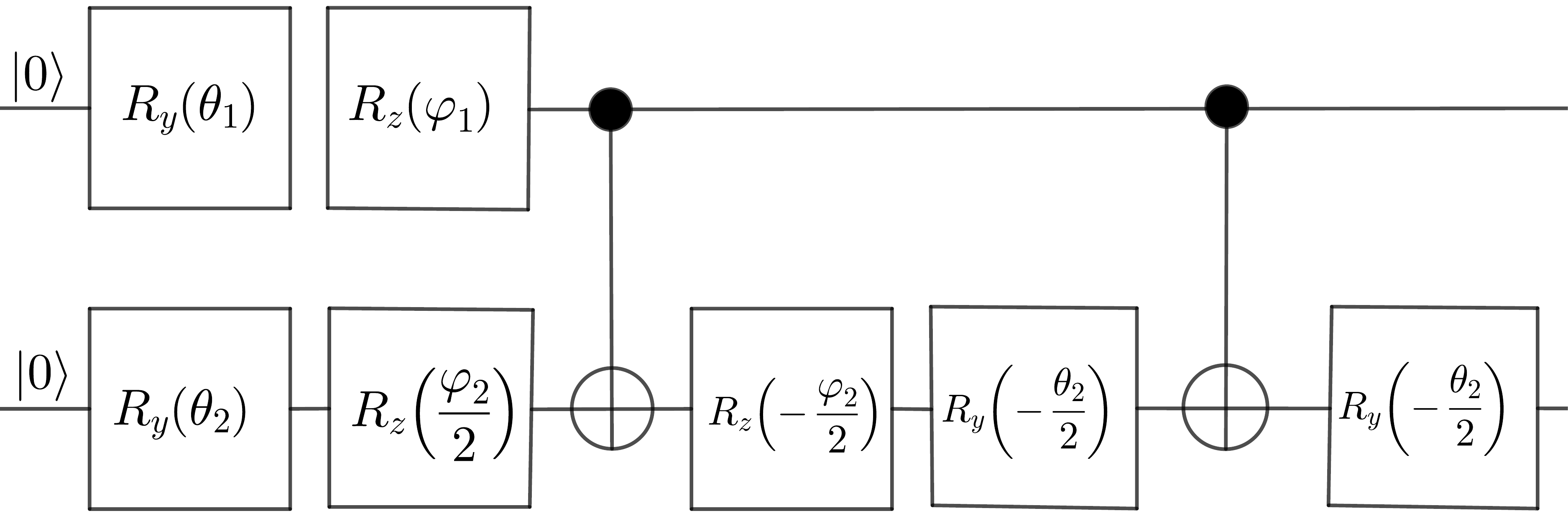}
\end{center}\begin{center}
(b) \includegraphics[width=0.8\hsize]{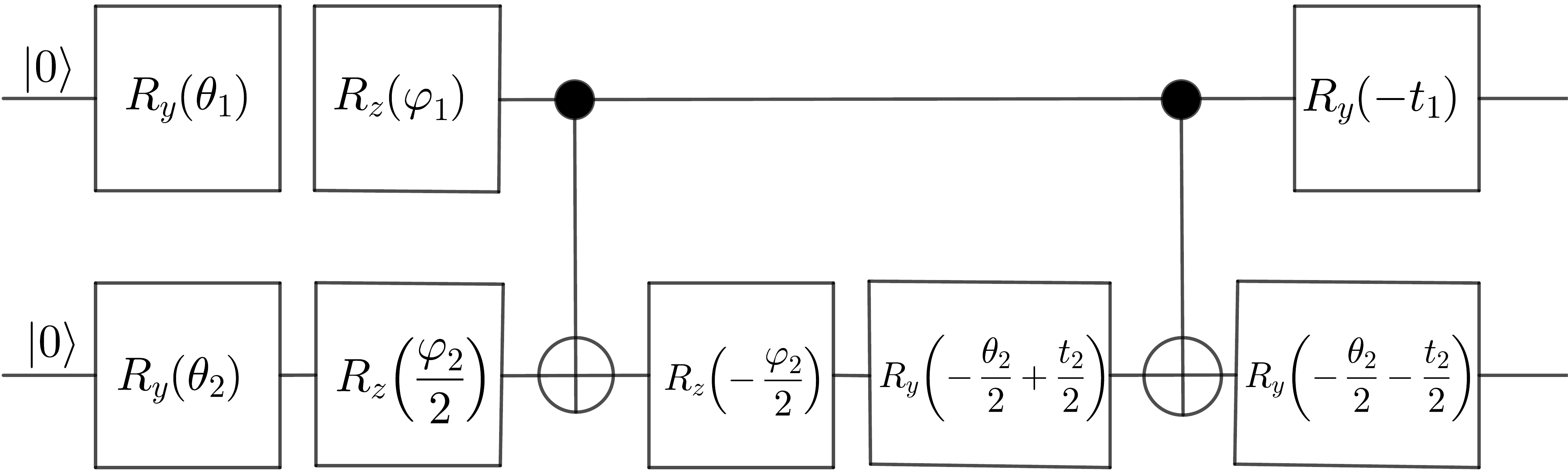}
\end{center}
\caption{Sorkin test circuits. (a)~Circuit to prepare arbitrary
  three-level states \eqref{sorkinst}: $R_\ell(\lambda)$ represents a
  rotation around the $\ell$ axis by an angle $\lambda$. (b)~Complete
  circuit that includes also the measurements. The parameters ${t}_1$
  and ${t}_2$ are used to select the different measurements. For
  example, the projection onto $|00\>+|10\>+|11\>$ discussed in the
  text is obtained by choosing ${t}_1=2\cos^{-1}\left(1/\sqrt{3}\right)$,
  ${t}_2=2\cos^{-1}\left(1/\sqrt{2}\right)$.}
    \label{f:sorkin}
\end{figure}

\begin{figure}
\includegraphics[width=.7\hsize]{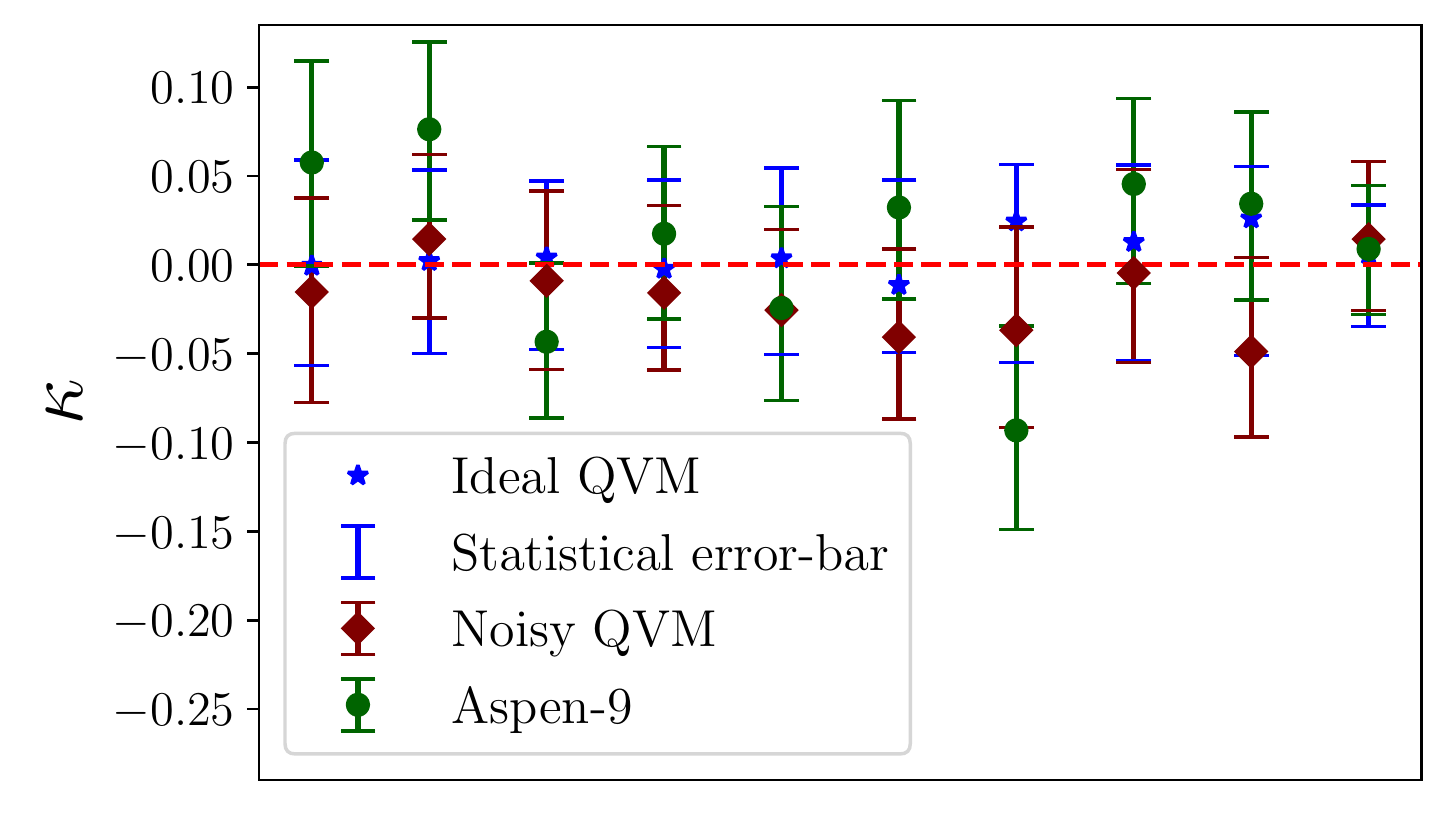}
\caption{Plot of $\kappa_3$ of Eq.~\eqref{kappa3} of the Sorkin test
  for randomly chosen states performed using the circuit of
  Fig.~\ref{f:sorkin}b. All obtained values are compatible with the
  theoretical value $\kappa_3=0$ expected from standard quantum
  mechanics. The error bars are produced using the same method as in
    the case of Peres' test above.}
  \label{f:res_sorkin}
\end{figure}

To implement Sorkin's test on a quantum computer we need to create an
arbitrary superposition of $n$ orthogonal pathways. This can be done
using $n$ qubits with a unary encoding \cite{simanraj} or, more
efficiently and in a less error-prone manner with $\log_2n$ qubits in
a binary encoding.  Start with $n=3$: the circuit to create arbitrary
three level states with binary encoding is presented in
Fig.~\ref{f:sorkin}a. It implements the transformation
$U(\theta_1, \varphi_1, \theta_2, \varphi_2)|00\>$ which prepares the
state
\begin{align}
\label{sorkinst}
  &\ket{\psi}= \cos\tfrac{\theta_1}{2} e^{-i \varphi_1/2} \ket{00} +
    \sin\tfrac{\theta_1}{2} e^{i \varphi_1/2} 
    \cos\tfrac{\theta_2}{2} e^{-i \varphi_2/2}\ket{10} \nonumber \\
  &+ \sin\tfrac{\theta_1}{2} e^{i \varphi_1/2} \sin\tfrac{\theta_2}{2} e^{i \varphi_2/2}\ket{11},
\end{align}
where $\theta_k$ and $\varphi_k$ are defined in the figure caption and
are chosen randomly.  To get the probabilities
\eqref{tre}-\eqref{probs}, we need to project $|\psi\>$ onto a state
$\<\psi'|$ such as $\left(\bra{00}+\bra{10}+\bra{11}\right)/\sqrt{3}$,
etc.  The state $\bra{\psi'}= \bra{00}U^\dag(t_1, f_1, t_2, f_2)$ is
implemented by the adjoint of the circuit of Fig.~\ref{f:sorkin}a with
appropriate $t_1,f_1,t_2,f_2$, and the projection is
$\braket{00|U^\dagger(t_1, f_1, t_2, f_2) U(\theta_1, \varphi_1,
  \theta_2, \varphi_2)|00}$. The simplified quantum circuit to
implement
$U^\dagger(t_1, 0, t_2, 0) U(\theta_1, \varphi_1, \theta_2,
\varphi_2)$ is shown in Fig.~\ref{f:sorkin}b.
Measurements in the computational basis can
be done by setting $t_1 = t_2 = 0$ and getting all the projections
from the same run of the circuit. Using this
circuit, we performed the three-level Sorkin test on a number of
randomly chosen states. The results are presented in
Fig.~\ref{f:res_sorkin} and they confirm that $\kappa_3$ is
statistically compatible with zero, as expected.

The extension to arbitrary $n$ can be obtained from the measurement of
the $n$-path probability: it can be implemented using a Hadamard gate
on each path followed by computational basis measurement if $n$ is a
power of two or, in general, from a circuit whose adjoint creates a
uniform superposition starting from a $\ket{0\cdots0}$ state. The
probability of obtaining all zeros from this measurement gives the
first term of the hierarchy, namely the first sum in \eqref{kappa_n},
i.e.~Eq.~\eqref{tre} for $n=3$. Then we only require 2-path and 1-path
probabilities that can be obtained with a trivial extension of the
above procedure: translate to binary and use two-qubit correlations
for the 2-path probabilities or measure the computational basis for
the 1-path probabilities. This is sufficient, in principle, to
incrementally scale the Sorkin test to large $n=2^N$ using $N$ qubits
\cite{simanraj}, although in practice, current NISQ device limitations
prevent us from testing for large $n$ as errors increase with increasing number of qubits.

\begin{figure}
    \centering
    \includegraphics[width=.7\hsize]{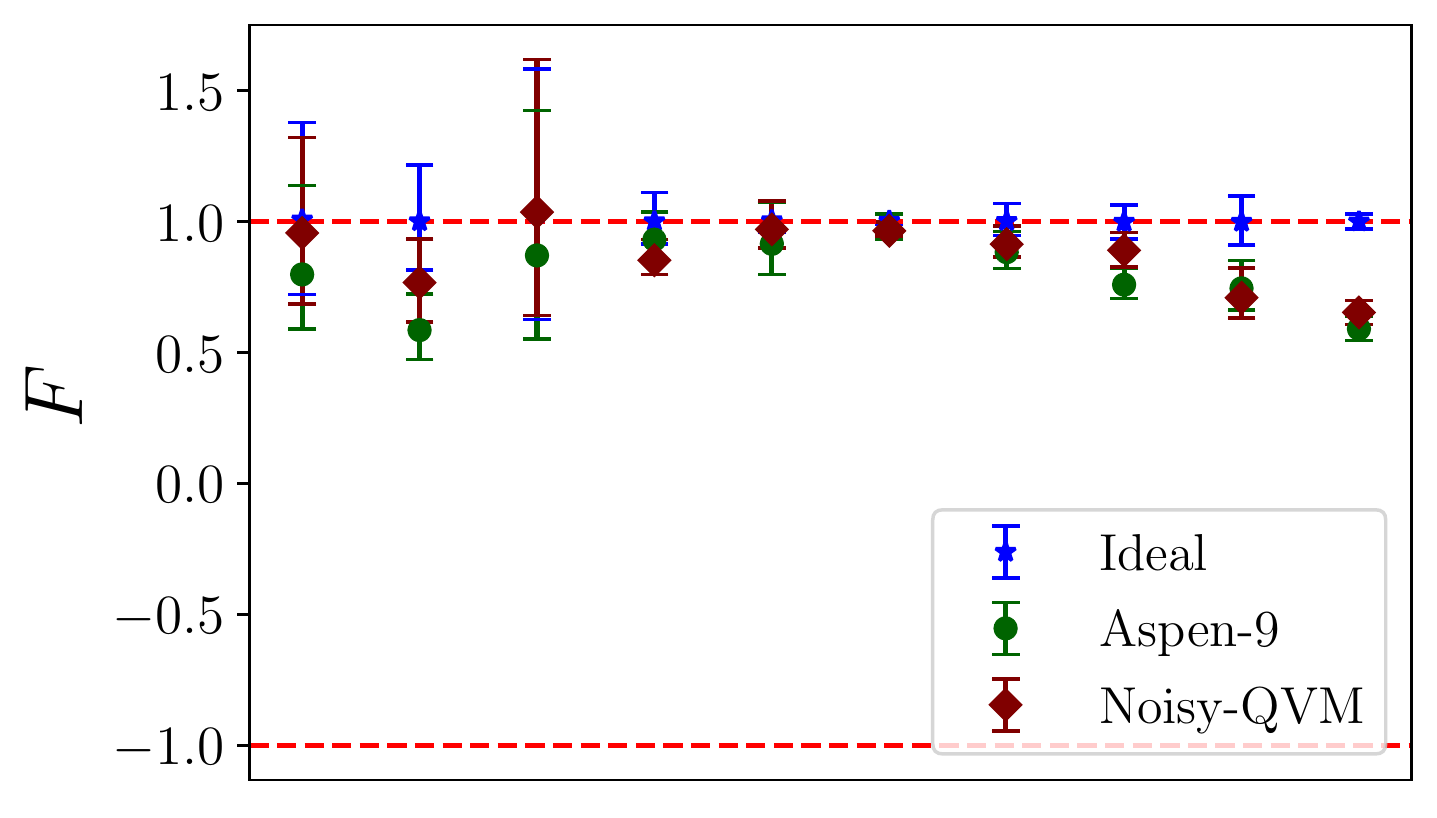}
    \includegraphics[width=.7\hsize]{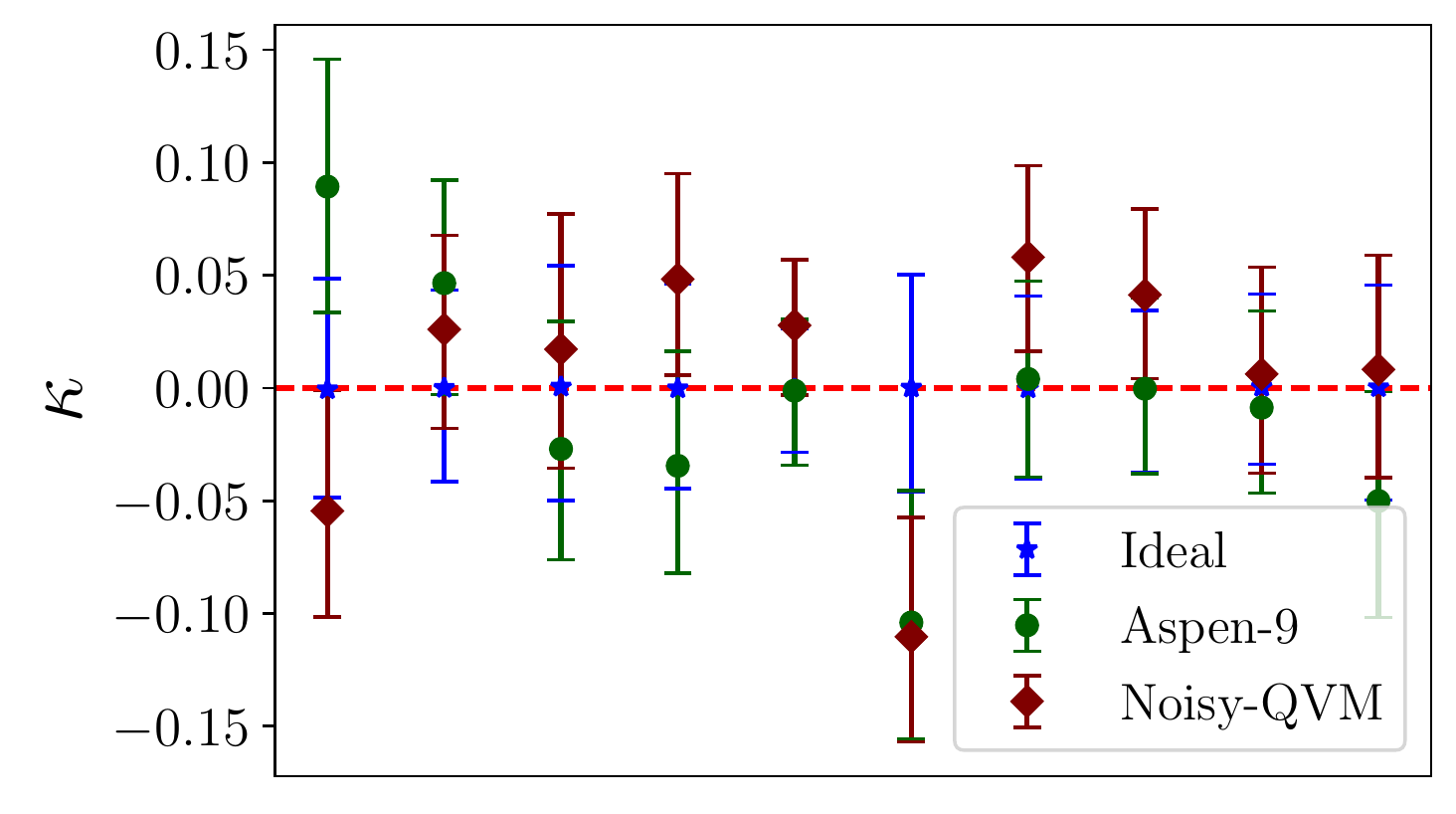}
    \caption{Results of Peres' and Sorkin's test using the same set of states and circuits for both. The data acquired to perform Sorkin's test in this case were sufficient to perform Peres' test and therefore, the same data-set is used to plot both $\kappa_3$ and $F$. The results of the joint Peres-Sorkin test are consistent with theoretical expectations, taking into account the readout and dephasing errors present in the system. These results are also consistent with the results of the standalone versions of the Peres' and Sorkin's test shown above (the values are, however, different because the set of random states used in each case is different).}
    \label{fig:peres-sorkin-joint}
\end{figure}

Interestingly, our algorithmic procedure allows us to perform the
Peres and Sorkin tests {\em jointly} for a class of states. Instead of
using the above procedure to prepare the Sorkin test state, we use the
Peres test circuit of Fig.~\ref{f:peres_circ} to produce and project a
set of randomly generated states of the form
$\ket{\psi_1\psi_2}=a\ket{00}+b\ket{01}+c\ket{10}+d\ket{11}$.
We then consider the projections onto
  $\ket{00}+\ket{01}+\ket{10}$, onto the two path-states
  $\ket{00}+\ket{01}$, $\ket{00}+\ket{10}$ and $\ket{01}+\ket{10}$ and
  onto the singles. For Sorkin test, we make an additional projection
  on $\ket{00} + \ket{01} + \ket{10}$ by replacing the CNOT and single
  qubit rotation at the end with the adjoint of the circuit that
  transforms
  $\ket{00} \rightarrow \left(\ket{00} + \ket{01} +
    \ket{10}\right)/\sqrt{3}$ and then measuring $\ket{00}$. Since the state $\ket{11}$ never appears in these
measurements, this procedure is equivalent to first projecting the
state $\ket{\psi_1\psi_2}$ onto the subspace spanned by $\ket{00}$,
$\ket{01}$ and $\ket{10}$ and then performing the Sorkin test on the
projected state (Results in
Fig.~\ref{fig:peres-sorkin-joint}).

\section{Conclusions}
We propose and implement quantum algorithms to test some of the
physical principles behind two postulates of quantum mechanics: the
complex nature of quantum Hilbert spaces and the form of the Born
rule.  We also perform both tests at the same time. We
present results on a NISQ device that can be
interpreted as a deep quantum benchmark of such devices.  We initially
believed that we could translate the Peres and Sorkin tests into
quantum algorithms in a straightforward manner, but we found we had to
modify these tests in a nontrivial way both due to the practical
limitations of current NISQ devices and to the fundamental limitations
of the gate model of quantum computation, which would require, as an
input to the un-modified Peres algorithm, the same quantity $\gamma$
that one then measures \cite{simanraj}.

\section{Acknowledgments}
LM acknowledges funding from the MIUR Dipartimenti di Eccellenza
2018-2022 project F11I18000680001 and from the U.S.  Department of
Energy, Office of Science, National Quantum Information Science
Research Centers, Superconducting Quantum Materials and Systems Center
(SQMS). We acknowledge support from Rigetti, and in particular from
Matt Reagor. US acknowledges partial support provided by the
Ministry of Electronics and Information Technology
(MeitY), Government of India under grant for Centre
for Excellence in Quantum Technologies with Ref. No.
4(7)/2020 - ITEA, QuEST-DST project Q-97 of the Govt. of India and the QuEST-ISRO research grant.


\begin{references}
\bibitem{feynman}R. P. Feynman, Simulating Physics with Computers,
  Int. J. Theor. Phys. {\bf 21}, 467 (1982).
\bibitem{woottersreal}W.K. Wootters, Optimal Information
  Transfer and Real-Vector-Space Quantum Theory. In: Chiribella G.,
  Spekkens R. (eds) Quantum Theory: Informational Foundations and
  Foils. Fundamental Theories of Physics, vol 181. Springer   (2016), 
arXiv:1301.2018.
\bibitem{stu}E.C.G.  Stueckelberg, Quantum Theory in Real Hilbert
  Space, Helv. Phys. Acta {\bf 33}, 727 (1960).
\bibitem{perescompl}A. Peres, Proposed Test for Complex versus
  Quaternion Quantum Theory, Phys. Rev. Lett. {\bf 42}, 683 (1979).
\bibitem{toni}M.-O. Renou, D. Trillo, M. Weilenmann, L.P. Thinh, A.
  Tavakoli, N. Gisin, A. Acin, M.  Navascues, Quantum physics needs
  complex numbers, arXiv:2101.10873 (2021).
\bibitem{walther} L.M. Procopio, L.A. Rozema, Z.J. Wong, D.R. Hamel, K. O'Brien, 
X. Zhang, B. Dakic, P. Walther, Single-photon test of hyper-complex quantum theories
  using a metamaterial, Nature Comm. {\bf 8} 15044 (2017). 
\bibitem{greg}R.  Keil, T. Kaufmann, T.  Kauten, S. Gstir, C.  Dittel,
  R. Heilmann, A. Szameit, G. Weihs, Hybrid waveguide-bulk multi-path
  interferometer with switchable amplitude and phase, APL Phot. {\bf
    1}, 081302 (2016).
\bibitem{gregor}S. Gstir, E. Chan, T. Eichelkraut, A. Szameit, R.
  Keil, G. Weihs, Towards probing for hypercomplex quantum mechanics
  in a waveguide interferometer, New. J. Phys. to be published,
  arXiv:2104.11577 (2021).
\bibitem{janwei}M.-C. Chen, et al., Ruling out real-number description
  of quantum mechanics, arXiv:2103.08123 (2021).
\bibitem{imagin}K.-D. Wu, T. Varun Kondra, S. Rana, C.  M. Scandolo,
  G.-Y. Xiang, C.-F. Li, G.-C. Guo A. Streltsov, Operational Resource
  Theory of Imaginarity, Phys. Rev. Lett. 126, 090401 (2021).
\bibitem{paoloreconstruction}G. Chiribella, G.M. D'Ariano, P.
  Perinotti, Informational derivation of quantum theory, Phys. Rev. A
  {\bf 84}, 012311 (2011).
\bibitem{sorkin}R.D. Sorkin, Quantum Mechanics as Quantum Measure
  Theory, Mod. Phys. Lett. A {\bf 9}, 3119 (1994).
\bibitem{caslav}B. Daki, T. Paterek, C. Brukner, Density cubes and
  higher-order interference theories, New J. Phys. {\bf 16}(2), 023028
  (2014).
\bibitem{zykowski} K. Zyczkowski, Quartic quantum theory: an extension
  of the standard quantum mechanics, J. Phys.  A 41, 355302 (2008).
\bibitem{lee}C.M. Lee, J.H. Selby, Higher-Order Interference in
  Extensions of Quantum Theory, Found. Phys. {\bf 47}, 
\bibitem{urbasi}U. Sinha, C. Couteau, T. Jennewein, R. Laflamme, G.
  Weihs, Ruling out multi-order interference in quantum mechanics,
  Science {\bf 329}, 418 (2010).
\bibitem{laflamme}D. K. Park, O. Moussa, R. Laflamme, Three path
  interference using nuclear magnetic resonance: a test of the
  consistency of Born's rule, New J. Phys. {\bf 14}, 113025 (2012).
\bibitem{werner}H. Kaiser, E. A. George, S. A. Werner, Neutron
  interferometric search for quaternions in quantum mechanics, Phys.
  Rev. A 29, 2276(R) (1984).
\bibitem{adler} S. L. Adler, Generalized quantum dynamics, Nucl. Phys.
  B {\bf 415}, 195 (1994).
\bibitem{octonions}S. De Leo, K. Abdel-Khalek, Octionionic Quantum
  Mechanics and Complex Geometry, Prog. Theor. Phys. {\bf 96}, 823
  (1996).
\bibitem{simanraj}S. Sadana, L. Maccone, U. Sinha, Quantum
  Computational tests of the Born Rule, in preparation.
\bibitem{rigettiweb} Rigetti SDK at
  \url{https://qcs.rigetti.com/sdk-downloads}. 
\bibitem{sourcecode}\url{http://www.rri.res.in/QuicLab/Peres_Sorkin/}
\bibitem{urbasi2}R. Sawant, J. Samuel, A. Sinha, S. Sinha, U. Sinha,
  Nonclassical Paths in Quantum Interference Experiments, Phys. Rev.
  Lett. {\bf 113}, 120406 (2014).
\bibitem{rengaraj}G. Rengaraj, U. Prathwiraj, S. N. Sahoo, R. Somashekhar, U. Sinha, New J. Phys. {\bf 20}, 063049 (2018).
\bibitem{usinha3}A. Sinha, A. H. Vijay, U. Sinha, Scientific Reports, {\bf 5}, 1--9 (2015).
\bibitem{gsch}I. S\"ollner, B. Gsch\"osser, P. Mai, B.  Pressl, Z.
  V\"or\"os, G. Weihs, Testing Born's rule in quantum mechanics for
  three mutually exclusive events, Found. Phys. {\bf 42}, 742 (2012).
\bibitem{luders}G.  L\"uders, \"Uber die Zustands\"anderung durch
  den Messprozess, Ann. Phys., Lpz. {\bf 8}, 322 (1951).
\end{references}
\end{document}